# Nanoscale Magnonic Neurons

K. G. Fripp, A. V. Shytov and V. V. Kruglyak

University of Exeter, Stocker Road, Exeter, EX4 4QL, United Kingdom

We use micromagnetic simulations to demonstrate neuron functionality of two-dimensional (2D) chiral magnonic resonators. Our design exploits nonlinear resonant scattering of spin waves propagating in a YIG medium from an edge mode of a permalloy nano-element. The reduced frequency and volume of the edge mode facilitate matching it to the YIG modes and give rise to their wide-angle scattering. As the amplitudes of the incident spin waves increase, the edge mode exhibits a positive nonlinear frequency shift. This shift leads to a complex frequency-dependent nonlinear variation of the amplitude and phase of spin waves scattered in different directions. We show that the scattered waves are strong enough to "activate" secondary neurons. This provides the connectivity required for combining our proposed neurons into 2D magnonic neural networks.

The rising significance of research in machine learning and artificial neural networks for modern society was recently recognized by the award of the 2024 Nobel Prize in Physics. Yet, the advent of the artificial intelligence (AI) brings about not only benefits but also the challenge of mounting energy costs of implementing AI with conventional, CMOS-based computers. Magnonics [1] might offer a viable alternative hardware platform for energy-efficient neuromorphic and, more generally, unconventional computing. This would use spin waves and their quanta (magnons) [2,3]) as information carriers that feature strong nonlinearity, modest propagation loss, and sub-micrometre wavelengths at GHz frequencies [1-5]. These properties have been used to build complex nonlinear reservoirs underpinning various forms of magnonic reservoir computing [4-8], showing different levels of trainability and success in common benchmark tasks.

An alternative, bottom-up approach involves devising a magnonic neuron first and then using it as a building block of more complex artificial neural networks. In Ref. [9], we explored nonlinear properties of chiral magnonic resonators [10-12] in view of their potential applications as artificial "neurons". However, only a one-dimensional (1D) system was considered under assumption of the incident and transmitted spin wave signals propagating along the same direction. This would limit options for connecting the neuron to those in subsequent layers. Moreover, the need to match the frequency of the propagating modes to those confined in the resonator led us to using the same material (Permalloy) for both the resonator and the magnonic medium. Yet, construction of large neural networks with a competitive computing efficiency dictates the use of low-damping materials, e.g. yttrium-iron garnet (YIG) [13,14], for the magnonic medium [15,16].

In this Letter, we use micromagnetic simulations to demonstrate neuron functionality of two-dimensional (2D) chiral magnonic resonators. Such neurons exploit nonlinear resonant scattering of spin waves propagating in a YIG medium from an edge mode of a permalloy disk nano-element. The reduced frequency and volume of the edge mode [17,18] facilitate matching it to the YIG modes and give rise to their wide-angle scattering. As the amplitudes of the incident spin waves increase, the edge mode exhibits a positive nonlinear frequency shift. This shift leads to a complex frequency-dependent nonlinear variation of the amplitude and phase of spin waves scattered in different directions. We show that the scattered waves have amplitude sufficient to "activate" secondary neurons. This provides the connectivity required for combining our proposed neurons into 2D magnonic neural networks [5,12].

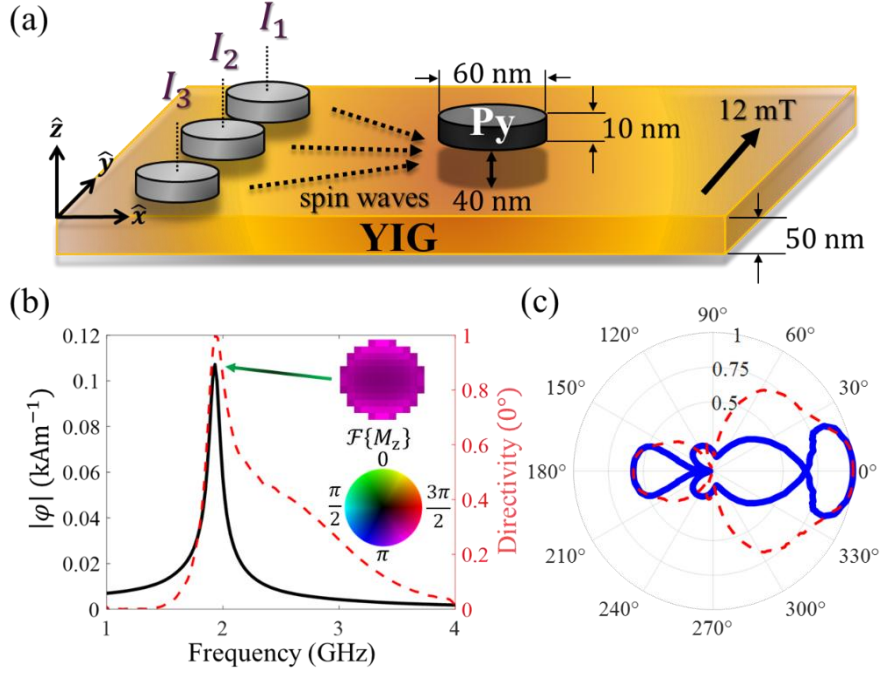

Figure 1. (a) Design of the magnetic neuron. Spin waves from sources $I_1$, $I_2$, and $I_3$ are scattered by a chiral magnonic resonator (the magnonic neuron) – a 10 nm thick permalloy (Py) [19] disk of 60 nm diameter spaced by 40 nm from a 50 nm thick YIG film [20]. A uniform bias magnetic field of 12 mT is applied in the $+y$ direction. (b) The spectra of the precessional response of the disk to a uniform linear excitation (solid black curve) and the normalised scattering directivity (dashed red curve) are shown for frequencies in the vicinity of the disk's lowest frequency resonance (edge mode). The inset shows the mode's profile, with the brightness and colour representing the Fourier amplitude and phase calculated from the out-of-plane component of the dynamic magnetisation. (c) The normalised directivity of the resonant scattering of spin waves at the frequency of the edge mode (1.935 GHz) is shown in the reciprocal space (red curve) and in the real space (blue curve), with the scattering angle of 0° corresponding to the $+x$ direction.

We envisage implementation of a magnonic neuromorphic device architecture based on a 'unit cell' building block shown in Fig.1(a). Three inputs $I_1$, $I_2$, and $I_3$ emit spin waves that are incident upon a permalloy disk that serves as a magnonic neuron. The neuron receives, nonlinearly transforms, and re-emits spin waves propagating linearly in a low-damping magnonic medium, such as YIG film. To show that such a system is suitable for performing neuromorphic computations, we characterise its response using finite-difference time-domain micromagnetic simulations with the help of the MuMax3 software [21]. The system is discretised using cuboidal cells with edge length of 5×5×10 nm³ on a grid of 2048×1024×10 cells $(x \times y \times z)$. 'Edge-smoothing' of order 4 is adopted to approximately correct the discretisation error at the edges. The mode spectrum of magnetic nanoelements (and especially the edge mode's frequency) is very sensitive both to the edge quality experimentally [22-24] and to the discretisation cell size numerically [25,26]. So, no attempt is made to elaborate our micromagnetic model beyond a qualitative agreement with existing experimental results [17,22-24]. To approximate an infinite extent of the film, periodic boundary conditions are applied in the in-plane dimensions, alone with a gradual increase of the Gilbert damping in the $\pm x$ directions. The corresponding procedures are described in the Supplementary material [27]. The boundaries between the regions of increased and uniform damping lie approximately 2 µm from the edges of the simulation grid along $x$. The system is relaxed to its static configuration in the bias magnetic field of 12 mT before applying a suitable dynamic magnetic field to study its linear and nonlinear dynamics.

The linear spectrum of the system is shown in Fig. 1 (b). The spectrum is obtained by exciting the region of the disk (but not the YIG film) by a weak, spatially asymmetric broadband field pulse and then summing the Fourier amplitudes calculated cell-wise from the disk's magnetization [17,18,22-26]. The lowest frequency resonance at 1.935 GHz lies well within the dipolar spin-wave continuum in YIG and is also the strongest and well separated from higher order modes. It corresponds to the disk's edge mode, whose frequency is reduced (compared to the Kittel frequency of a plane film made of the same material) by the demagnetising field due to the magnetic charges formed at the disk edges [17,18].

The dynamic dipolar coupling of the spin waves propagating in YIG to the edge mode (the 'local' mode) leads to their resonant scattering. To obtain its directivity shown in Fig. 1 (c), a 1.935 GHz frequency plane spin wave is launched in the $+x$ direction (0°), after which the scattered wave's Fourier amplitudes are calculated from the YIG film's magnetisation for different reciprocal space directions [28]. The directivity reveals that the resonant scattering is chiral, i.e. the disk acts as a chiral magnonic resonator with minimal backscattering [12]. The strong forward-backward asymmetry facilitates implementing feed-forward networks of such neurons. In the reciprocal space, the scattering occurs into a wide angle of about ±60°. However, the anisotropic magnonic dispersion of the in-plane magnetised YIG film [29,30] means that the scattering angle shrinks by about a half in the real space, i.e. when considered in terms of the directions of the group velocity. This scattering angle determines the number of other neurons with which the given one would be able to communicate (i.e. receive from or feed to) the signal.

Before we present our results of nonlinear spin-wave scattering from the nano-disk, we first characterise the nonlinearity of its edge mode resonance. We drive the nano-disk placed above the YIG film (but not the film itself) by a monochromatic spatially-uniform circularly polarised magnetic field of the form

$$B_{\text{ex}}(t) = B_0\big(\hat{x}\sin(2\pi ft) + \hat{z}\cos(2\pi ft)\big), \tag{1}$$

where $B_0$ and $f$ are the driving field's amplitude and frequency, respectively. The circular polarisation mimics that of the dynamic stray field of propagating plane spin waves with a positive wave number [31]. The disk's response to this field therefore provides a reference when considering nonlinear spin-wave scattering later. Given that the spin-wave wavelength $\lambda$ is at least an order of magnitude longer than the diameter of the disk, equation (1) provides a good approximation of the spatial variation of the dynamic stray field of incident spin waves in the region of the disk. To speed up the simulations, the grid size is reduced to 320×320×10 cells in the ($x \times y \times z$) directions with the cell size of 5×5×10 nm³, while increasing the number of grid images in the periodic boundary conditions and adding absorbing boundary conditions in the $\pm y$ directions [27]. When the excitation strength is small, e.g. for $B_0 = 0.005\ \text{mT} \equiv B_{\text{linear}}$, the response of the edge mode is linear and can be modelled as a Lorentzian in terms of the amplitude $|\varphi|$ of the local mode

$$[(\omega - \omega_0)^2 + \Gamma_{\text{tot}}^2] \cdot |\varphi|^2 = A_{\text{linear}}^2, \tag{2}$$

where $\omega_0$ is the mode's angular frequency, $\Gamma_{\text{tot}}$ is the resonance linewidth that includes contributions from both Gilbert damping and radiative damping [11], and $A_{\text{linear}}$ is a product of $B_{\text{linear}}$ and a coefficient describing the mode's coupling strength to the field given by Eq. (1). We fit Eq. (2) to the edge mode's response obtained from micromagnetic simulations as

$$|\varphi_{\text{data}}(B_0, f)| = \frac{2}{N_{\text{FFT}}} \cdot \sum_{i=1} |\mathcal{F}\{M_x(i)\}|, \tag{3}$$

where $\mathcal{F}\{M_x(i)\}$ is the fast Fourier transform in time of the $x$ component of the dynamic magnetisation for the $i^{\text{th}}$ cell, with index $i$ enumerating all cells in the disk with non-zero magnetisation. The pre-factor of $\frac{2}{N_{\text{FFT}}}$ (where $N_{\text{FFT}}$ is the length of the data in the time domain) ensures normalisation of the Fourier transformed real-valued data. So, $|\varphi_{\text{data}}(B_0, f)|$ has the meaning of the $M_x$ oscillation amplitude averaged over the disk's volume. The fit yields $\omega_0 = 2\pi \times 1.935\ \text{GHz}$, $A_{\text{linear}} = 2\pi \times 401.21\ \text{Am}^{-1}\text{GHz}$, and $\Gamma_{\text{tot}} = 2\pi \times 0.077\ \text{GHz}$, which we use below.

To characterise the nonlinearity of the nano-disk, we excite it with microwave fields of increased amplitude $B_0$. The nonlinearity parameter $\tilde{\lambda}$ is then extracted by fitting the simulated results to Eq. (2) with a quadratic frequency shift $\tilde{\lambda}|\varphi|^2$ added [11], i.e.

$$[(\omega - \omega_0 + \tilde{\lambda}|\varphi|^2)^2 + \Gamma_{\text{tot}}^2] \cdot |\varphi|^2 = \left(\frac{B_0}{B_{\text{linear}}}\right)^2 A_{\text{linear}}^2. \tag{4}$$

The fit yields $\tilde{\lambda} = -7.229 \times 10^{-13}$ GHz A$^{-2}$ m$^2$, with the fitted curves shown in Fig. 2 (a) and the magnitude of nonlinear shift of the edge mode's resonance frequency shown as a function of $B_0$ in Fig. 2 (b).

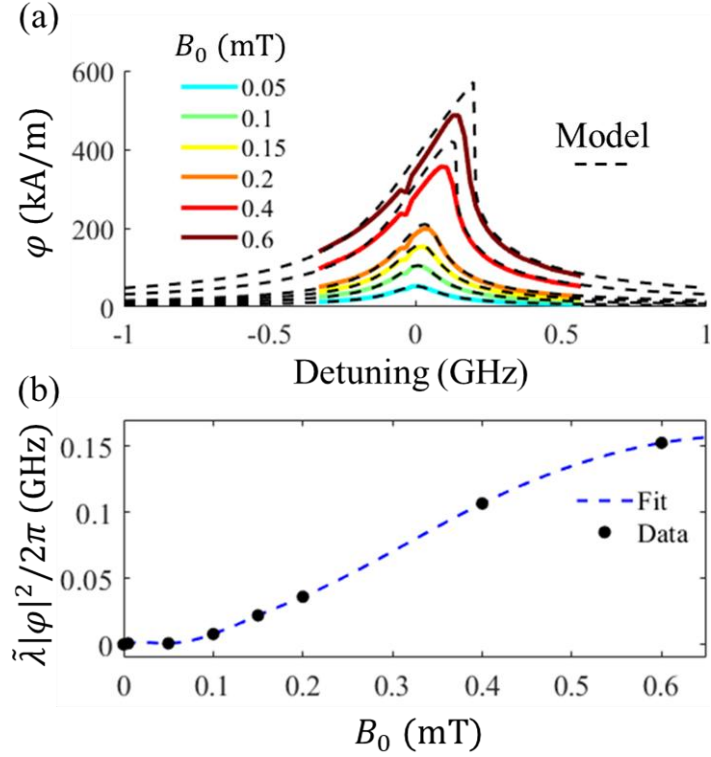

Figure 2. (a) The local mode amplitude $|\varphi|$ is shown as a function of the frequency detuning, $f - f_0$. The coloured solid lines show the data obtained from micromagnetic simulations for different values of the amplitude $B_0$ of the driving field. The black dashed lines show their fits to Eq. (4). (b) The nonlinear frequency shift $\tilde{\lambda}|\varphi|^2/2\pi$ obtained from micromagnetic simulations (black circles) is shown as a function of $B_0$. The blue dashed line is a cubic spline fit to the data.

Figure 2 shows that the sign of the nonlinearity is negative, so that the edge mode resonance is shifted towards higher frequencies ('blue shift'). This contrasts with our previous findings for a stripe-shaped chiral magnonic resonator [9], which showed a positive nonlinearity (red shift) for both the fundamental and first dark modes. Using the methodology described above, we obtain $\tilde{\lambda} = 8.653 \times 10^{-12}$ GHz A$^{-2}$ m$^2$ for the stripe-based system from Ref. [9] (see the Supplementary materials [27]). The origin of the sign difference of the nonlinearity lies in the nature of the edge mode. Its frequency is dominated by the effect of the static (rather than dynamic, as for the stripe's modes in Ref. [9]) internal field, i.e. the sum of the bias and demagnetising fields. Hence, as the amplitude of precession increases, this static demagnetising field decreases, and so, both the static internal field and the mode frequencies increase. The edge mode is confined in the 'pockets' of this internal field [17] and is therefore most sensitive to its change. The nonlinear model agrees well with the data up to the excitation amplitude $B_0 = 0.6$ mT (Fig. 2 (a)). At higher values of $B_0$, the slope of the curve decreases and then the sign of the nonlinearity reverses (not shown). This behaviour was not observed in Ref. [9] and is not captured by Eq.(4). Accounting for the presence of additional nonlinearities (likely explaining this behaviour) is beyond the scope of this paper.

The key attribute of the disk's resonant nonlinear response that makes it a viable magnonic neuron is the strong nonlinear frequency shift achieved at modest excitation field amplitudes ($B_0 \leq 0.6$ mT). Notably, when fields of such a strength are induced by spin waves propagating in the YIG medium (e.g. under the disk), the response of the YIG medium itself is still linear. Not only does this satisfy the assumption of the resonant chiral scattering model [9] but also minimises propagation losses due to nonlinear damping. This contrasts e.g. with Ref. [5] where two orders of magnitude stronger excitation fields were used.

Let us now consider the effect of the nonlinearity of the disk's edge mode on the scattering of incident spin waves. We adopt the geometry shown in Fig. 1 (a). The disk is placed at the location (–0.33 μm, 0 μm) with the origin defined in the middle of the simulation domain. The spatial dependence of the dynamic field in the three two-dimensional Gaussian spin-wave sources $I_1$, $I_2$, and $I_3$ is given by

$$I_n(x, y) = \exp\left(-\frac{(x-x_n)^2 + (y-y_n)^2}{2\sigma^2}\right), \tag{5}$$

where $n = 1,2,3$, $x_1 = x_2 = x_3 = -2.7$ μm, $y_1 = 0.6$ μm, $y_2 = 0$ μm, $y_3 = -0.6$ μm, and the value of $\sigma$ corresponds to the full width at half maximum (FWHM) of $0.35$ μm. Such a small FWHM ensures that spin waves are emitted into a wide angle. Each source is driven as described by Eq. (1) for the same excitation amplitude $B_0$. For each combination of $B_0$ and excitation frequency, we compute patterns of scattered spin waves and their stray dynamic magnetic field at a height of $40$ nm above the top surface of the YIG film. The patterns of the $x$ component of this field are presented in Fig. 3. Each pattern is normalised by the field's strength $B_\mathrm{I}$ obtained for the location of the disk in a reference simulation run under identical conditions but without the disk. The field patterns from the same reference simulations (Fig. 3 (a) and (e)) are also used to select the disk's location so that it corresponds to a region of enhanced dynamic magnetic field. This enhancement results from the constructive interference of spin waves emitted from the three point-like sources and forming caustic beams due to the anisotropy of the magnonic dispersion relation in the in-plane magnetised YIG medium [29,30].

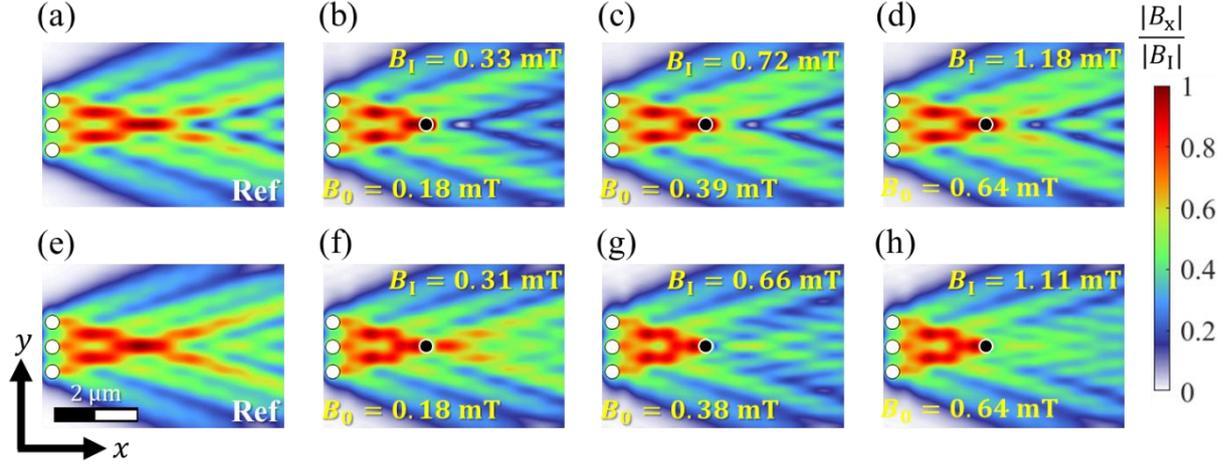

Figure 3. The patterns of the stray magnetic field, $|B_x|$, created by the scattered spin waves above the YIG medium and normalised by the reference magnetic field $B_I$ are shown for different excitation field amplitudes $B_0$. Panels (a) to (d) correspond to the spin-wave frequency equal to that of the edge mode (1.935 GHz), while panels (e) to (h) correspond to 2.08 GHz, i.e. just above the edge-mode resonance. Panels (a) and (e) exemplify reference patterns from which $B_I$ values are extracted for each frequency and excitation amplitude $B_0$. The positions of the spin-wave sources and the disk are indicated by the white and black circles, respectively.

Turning to simulations with the disk present, we look first at the results obtained with the lowest excitation amplitude of 0.18 mT for the spin-wave frequencies of 1.935 GHz (Fig. 3 (b)) and 2.08 GHz (Fig. 3 (f)). In contrast to the reference simulations, we see a stark difference between the patterns obtained for the two frequencies behind the disk within the 60° cone around the horizontal centre line. At this excitation strength, the spin waves' stray magnetic field is below 0.33 mT, for which the nonlinear frequency shift of the edge-mode resonance is significant yet smaller than the edge-mode's half-linewidth (Fig. 2). So, at 1.935 GHz (Fig. 3 (b)), the incident spin waves are still at resonance with the edge mode, and their scattering is strong and accompanied by a phase shift of around $\pi$ [9,11]. This phase shift leads to the destructive interference and associated suppression of the spin waves' stray field behind the disk. As the frequency increases to 2.08 GHz (Fig. 3 (f)), the strength of the resonant scattering is reduced and the phase shift deviates from $\pi$, weakening the field suppression. As we increase the excitation strength to 0.38 mT, the positive shift of the edge mode frequency (Fig. 2) tunes it into resonance with the incident spin waves at 2.08 GHz, leading to the suppression of spin-wave transmission observed in Fig. 3 (g). The resonance is frustrated again at the excitation amplitude of 0.64 mT, leading to the partial recovery of the transmission (Fig. 3 (h)). In comparison, the enhancement of the spin-wave transmission at 1.935 GHz as the excitation amplitude is increased to 0.38 mT (Fig. 3 (c)) and then 0.64 mT (Fig. 3 (d)) is relatively slow. This is due to the nonlinearity-induced asymmetry of the resonant line shape evident from Fig. 2 (a). Nonetheless, the transmission pattern at 0.64 mT (Fig. 3 (d)) is very similar to that when the disk is absent altogether (Fig. 3 (a)). That is, we reach the effective 'activation' state of our 'neuron', such that the spin wave 'signal' is transmitted almost unperturbed.

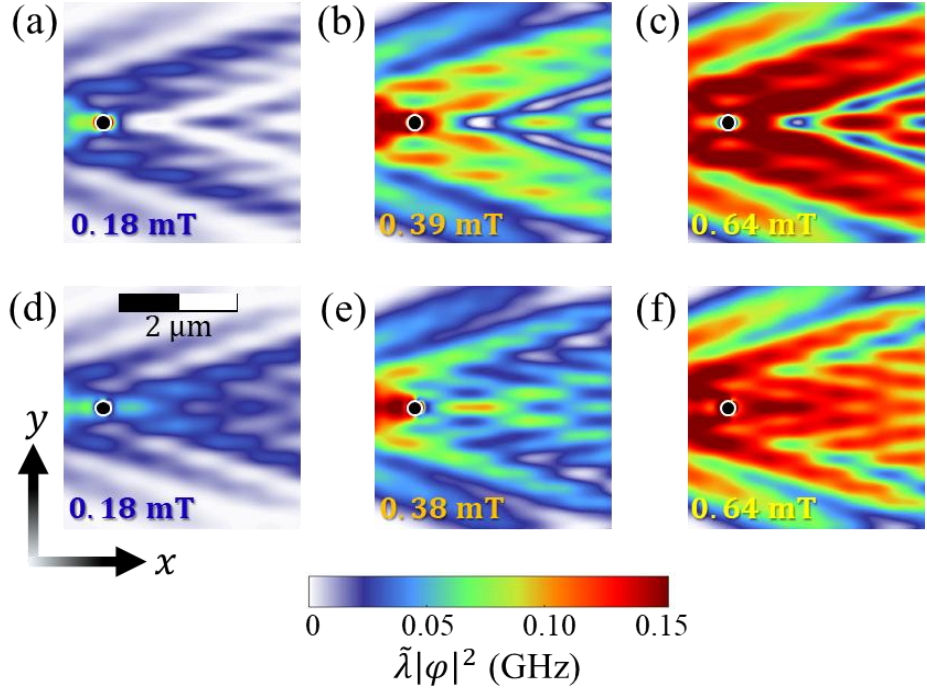

Figure 4. Spatial distributions of the nonlinear frequency shift $\tilde{\lambda}|\varphi|^2$ that would be induced in a secondary disk placed in the scattered wave of the primary disk. Panels (a)–(c) and (d)–(f) correspond to the spin-wave frequencies of 1.935 GHz and 2.08 GHz, respectively. In each panel, the position of the primary disk is indicated by a green circle, while the labels indicate the excitation amplitude of the spin-wave sources.

In a feed-forward neural network, each magnonic neuron must not only respond nonlinearly to the incident spin waves but must also be able to 'talk' to neurons in the network's next layer. This demands that the amplitude of the stray field induced by the scattered wave be strong enough to activate those secondary neurons. To illustrate how our system satisfies this requirement, we map the data from Fig. 3 into the "secondary frequency shift", i.e. the nonlinear frequency shift induced in a secondary neuron by the stray field produced in its position by the scattered spin waves (Fig. 4). This is done by equating the excitation amplitude from Fig. 2 (b) to the stray-field amplitude in Fig. 3. We see that at the excitation amplitude of 0.18 mT, the secondary frequency shift remains <25 MHz for 1.935 GHz (Fig. 4 (a)), while being only marginally greater for 2.08 GHz (Fig. 4 (d)). At the medium excitation amplitudes, the secondary frequency shift exceeds 100 MHz (Fig. 4 (b) and (e)), being especially strong along the caustic beams, another consequence of the anisotropic spin-wave dispersion [29,30]. At the highest excitation amplitude (Fig. 4 (c) and (f)), the scattered wave is strong enough to produce secondary frequency shifts of 150 MHz, with a wider angle and so, potentially a greater number of secondary neurons covered for the frequency of 1.935 GHz (Fig. 4 (c)). At both frequencies, the spatial nonuniformity of the scattered-wave patterns provides a range of locations where placement of secondary disks is possible. By placing disks in particular positions, we can 'weight' their responses to the wave scattered from the primary disk. Such a disk array would then act as a multi-level neural network designed to perform a specific function, e.g. a logic gate or a gate array.

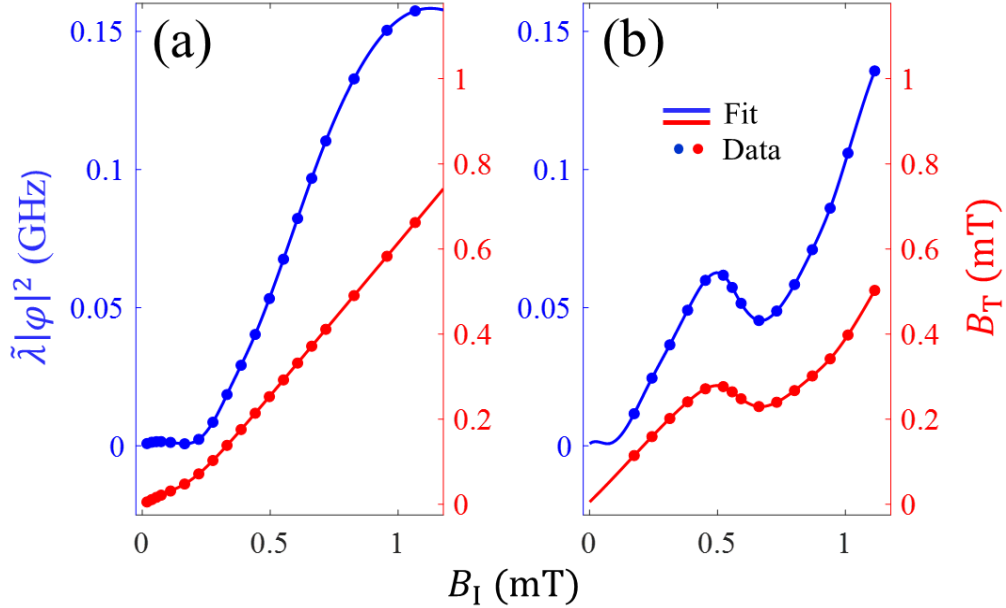

Figure 5. The secondary frequency shift $\tilde{\lambda}|\varphi|^2$ (left axis) and the scattered spin-wave stray field amplitude $B_T$ (right axis) at the locations of $(0.76\ \mu m, \pm 0.35\ \mu m)$ are shown for the frequencies of (a) 1.935 GHz and (b) 2.08 GHz, as a function of the amplitude $B_I$ of the incident spin waves. The solid circles show the data derived from micromagnetic simulations, whilst the solid lines are cubic spline fits of the data.

Figure 5 shows the stray magnetic field $B_T$ and corresponding secondary frequency shift produced by the scattered spin waves at a specific location of $(0.76\ \mu m, \pm 0.35\ \mu m)$. The curves show the familiar behaviour of activation and power-limiting [9] for spin waves at the resonance frequency (Fig. 5 (a)) and just above (Fig. 5 (b)). In Fig. 5 (a), we see a sigmoid-like curve where the secondary frequency shift plateaus around $B_I = 1$ mT. In Fig. 5 (b), the secondary frequency shift increases and then passes through a point of inflection at around $B_I = 0.5$ mT, before increasing again. The field reaches 50-60% of the level due to the incident spin waves, and the corresponding secondary frequency shift exceeds 0.1 GHz, suggesting a possibility of activating a hypothetical secondary neuron at the considered location and therefore of building artificial neural networks from magnonic proposed here. Such neural networks and their properties are however beyond the topic of this paper.

In summary, we have used micromagnetic simulations to demonstrate a two-dimensional chiral magnonic resonator as a magnonic neuron. Its functionality is based on the nonlinear resonant scattering of propagating spin waves from the edge spin-wave mode confined in a permalloy disk located above a low-damping YIG magnonic medium. Depending on the frequency and amplitude of the incident spin waves, we observe complex patterns of the scattered waves and show that their dynamic dipolar field could be strong enough to activate hypothetical magnonic neurons in the next layer of the network. Our results therefore suggest that the proposed magnonic neurons could serve as viable building blocks of future 2D artificial neural networks.

The research leading to these results has received funding from the UK Research and Innovation (UKRI) under the UK government's Horizon Europe funding guarantee (Grant No. 10039217) as part of the Horizon Europe (HORIZON-CL4-2021-DIGITAL-EMERGING-01) under Grant Agreement No. 101070347 (MANNGA project). Co-funded by the European Union. Views and opinions expressed are however those of the authors only and do not necessarily reflect those of the European Union or the European Health and Digital Executive Agency (HADEA). Neither the European Union nor the granting authority can be held responsible for them.